\input harvmac.tex


\def\unlockat{\catcode`\@=11}

\def\lockat{\catcode`\@=12}

\unlockat


\def\newsec#1{\global\advance\secno by1\message{(\the\secno. #1)}
\global\subsecno=0\global\subsubsecno=0
\global\deno=0\global\prono=0\global\teno=0\eqnres@t\noindent
{\bf\the\secno. #1} \writetoca{{\secsym}
{#1}}\par\nobreak\medskip\nobreak}
\global\newcount\subsecno \global\subsecno=0
\def\subsec#1{\global\advance\subsecno
by1\message{(\secsym\the\subsecno. #1)}
\ifnum\lastpenalty>9000\else\bigbreak\fi\global\subsubsecno=0
\global\deno=0\global\prono=0\global\teno=0
\noindent{\it\secsym\the\subsecno. #1} \writetoca{\string\quad
{\secsym\the\subsecno.} {#1}}
\par\nobreak\medskip\nobreak}
\global\newcount\subsubsecno \global\subsubsecno=0
\def\subsubsec#1{\global\advance\subsubsecno by1
\message{(\secsym\the\subsecno.\the\subsubsecno. #1)}
\ifnum\lastpenalty>9000\else\bigbreak\fi
\noindent\quad{\secsym\the\subsecno.\the\subsubsecno.}{#1}
\writetoca{\string\qquad{\secsym\the\subsecno.\the\subsubsecno.}{#1}}
\par\nobreak\medskip\nobreak}

\global\newcount\deno \global\deno=0
\def\de#1{\global\advance\deno by1
\message{(\bf Definition\quad\secsym\the\subsecno.\the\deno #1)}
\ifnum\lastpenalty>9000\else\bigbreak\fi \noindent{\bf
Definition\quad\secsym\the\subsecno.\the\deno}{#1}
\writetoca{\string\qquad{\secsym\the\subsecno.\the\deno}{#1}}}

\global\newcount\prono \global\prono=0
\def\pro#1{\global\advance\prono by1
\message{(\bf Proposition\quad\secsym\the\subsecno.\the\prono #1)}
\ifnum\lastpenalty>9000\else\bigbreak\fi \noindent{\bf
Proposition\quad\secsym\the\subsecno.\the\prono}{#1}
\writetoca{\string\qquad{\secsym\the\subsecno.\the\prono}{#1}}}

\global\newcount\teno \global\prono=0
\def\te#1{\global\advance\teno by1
\message{(\bf Theorem\quad\secsym\the\subsecno.\the\teno #1)}
\ifnum\lastpenalty>9000\else\bigbreak\fi \noindent{\bf
Theorem\quad\secsym\the\subsecno.\the\teno}{#1}
\writetoca{\string\qquad{\secsym\the\subsecno.\the\teno}{#1}}}
\def\subsubseclab#1{\DefWarn#1\xdef
#1{\noexpand\hyperref{}{subsubsection}%
{\secsym\the\subsecno.\the\subsubsecno}%
{\secsym\the\subsecno.\the\subsubsecno}}%
\writedef{#1\leftbracket#1}\wrlabeL{#1=#1}}

\lockat


\def\IB{\relax\hbox{$\inbar\kern-.3em{\rm B}$}}
\def\IC{\relax\hbox{$\inbar\kern-.3em{\rm C}$}}
\def\ID{\relax\hbox{$\inbar\kern-.3em{\rm D}$}}
\def\IE{\relax\hbox{$\inbar\kern-.3em{\rm E}$}}
\def\IF{\relax\hbox{$\inbar\kern-.3em{\rm F}$}}
\def\IG{\relax\hbox{$\inbar\kern-.3em{\rm G}$}}
\def\IGa{\relax\hbox{${\rm I}\kern-.18em\Gamma$}}
\def\IH{\relax{\rm I\kern-.18em H}}
\def\IK{\relax{\rm I\kern-.18em K}}
\def\IL{\relax{\rm I\kern-.18em L}}
\def\IP{\relax{\rm I\kern-.18em P}}
\def\IR{\relax{\rm I\kern-.18em R}}
\def\IZ{\relax\ifmmode\mathchoice
{\hbox{\cmss Z\kern-.4em Z}}{\hbox{\cmss Z\kern-.4em Z}}
{\lower.9pt\hbox{\cmsss Z\kern-.4em Z}} {\lower1.2pt\hbox{\cmsss
Z\kern-.4em Z}}\else{\cmss Z\kern-.4em Z}\fi}

\def\II{\relax{\rm I\kern-.18em I}}

\def\frac#1#2{{#1\over#2}}
\def\CA {{\cal A}}

\def\CF {{\cal F}}


\def\pr{\partial}



\def\inbar{\,\vrule height1.5ex width.4pt depth0pt}
\font\cmss=cmss10 \font\cmsss=cmss10 at 7pt


\font\manual=manfnt \def\dbend{\lower3.5pt\hbox{\manual\char127}}


\def\boxit#1{\vbox{\hrule\hbox{\vrule\kern8pt
\vbox{\hbox{\kern8pt}\hbox{\vbox{#1}}\hbox{\kern8pt}}
\kern8pt\vrule}\hrule}}
\def\mathboxit#1{\vbox{\hrule\hbox{\vrule\kern8pt\vbox{\kern8pt
\hbox{$\displaystyle #1$}\kern8pt}\kern8pt\vrule}\hrule}}

\Title{ \vbox{\baselineskip12pt \hbox{hep-th/0105245}
 \hbox{ITEP-TH-25/01} \hbox{YCTP-SS4-01 } \hbox{} }} {\vbox{
\centerline{On non-Abelian Structures}
\vskip 0.7cm
\centerline{in}
\vskip 0.7cm
\centerline{Field Theory of Open Strings}}}

\medskip \centerline{\bf Anton A. Gerasimov $^{1}$ and Samson L. Shatashvili
$^{2}$\footnote{*}{On leave of absence from St.
Petersburg Branch of Steklov Mathematical Institute,
 Fontanka,
St. Petersburg, Russia.}}

\vskip 0.5cm \centerline{\it $^{1}$ Institute for Theoretical and
Experimental Physics, Moscow, 117259, Russia} \centerline{\it
$^{2}$ Department of Physics, Yale University, New Haven, CT
06520-8120 }
\vskip 1cm Multi-brane backgrounds are studied in the framework of
 the background independent open string field
theory. A simple description of the non-abelian degrees of freedom
is given.
 Algebra of the differential operators acting on the space of
 functions on the space-time
 provides a natural tool for the discussion of this phenomena.

\medskip
\noindent

\Date{May 19, 2001}

\newsec{Introduction}

The understanding of the relation between (large $N$) gauge
theories and string theories \ref\thooft{G. 't Hooft, "A Planar
Diagram Theory For Strong Interactions," Nucl. Phys. {\bf B72}
(1974) 461.}, \ref\Wilson{K. Wilson, Phys. Rev. {\bf D10} (1974) 2445.},
\ref\KogSus{J. Kogut and L. Susskind, Phys. Rev. {\bf D11} (1975) 395.},
\ref\Pol{ A. Polyakov,  Nucl. Phys. {\bf B164} (1979) 171.}
(see also discussion in \ref\Ployakov{A. M.
Polyakov, "String Theory And Quark Confinement," hep-th/9711002,
Nucl. Phys. Proc. Suppl. {\bf 68} (1998) 1.}),
 is probably one of the most
important problems in the modern theoretical physics. During the
last decade this connection showed up in string theory literature
with various faces: multiple $D$-branes backgrounds \ref\Dbr{J.
Polchinski, "Dirichlet-Branes And Ramond-Ramond Charges,"
hep-th/9510017, Phys. Rev. Lett. {\bf 75} (1995) 4724; E. Witten,
 "Bound States Of Strings And p-branes," hep-th/9510135,  Nucl. Phys. {\bf
B460} (1996) 335.}, Matrix Theory proposal \ref\BFSS{T. Banks, W.
Fischler, S.  H. Shenker and L. Susskind, "M Theory As A Matrix
Model: A Conjecture," hep-th/9610043, Phys. Rev. {\bf D55} (1997)
5112.}, AdS/CFT correspondence \ref\Mald{J. Maldacena, "The Large
$N$ Limit Of Superconformal Field Theories And Supergravity",
hep-th/9711200,  Adv. Theor. Math. Phys. {\bf 2} (1998) 231; Int.
J. Theor. Phys. {\bf 38} (1999) 1113.}, \ref\GKP{S. S. Gubser, I.
R. Klebanov and A. M. Polyakov, "Gauge Theory Correlators From
Noncritical String Theory," hep-th/9802109, Phys. Lett. {\bf B428}
(1998) 105.}, \ref\Witttwo{E. Witten, "Anti-de Sitter Space And
Holography," hep-th/9802150,  Adv. Theor. Math. Phys. {\bf 2}
 (1998) 253.}, various constructions of the solitonic objects in
terms of matrix degrees of freedom \ref\Solitone{R. Gopakumar, S.
Minwalla and A. Strominger, "Noncommutative Solitons,"
hep-th/003160,  JHEP {\bf 0005:020} (2000).}, \ref\Solittwo{K. Dasgupta, S. Mukhi
and G. Rajesh, "Noncommutative Tachyons," hep-th/0005006,  JHEP
{\bf 0006:022} (2000).},  \ref\Solitthree{
Jeffrey A. Harvey, Per Kraus, Finn Larsen and
Emil J. Martinec, {\it ``D-branes and Strings as
Non-commutative Solitons"}, hep-th/0005031, JHEP 0007 (2000) 042;
Jeffrey A. Harvey, Per Kraus and Finn Larsen,
{\it "Exact Noncommutative Solitons",} hep-th/0010060,
JHEP 0012 (2000) 024.},
\ref\Solitfour{D. J. Gross and N. A. Nekrasov, "Monopoles
And Strings In Noncommutative Gauge Theory," hep-th/0005204,  JHEP
{\bf 0007:034} (2000); D. J. Gross and N. A. Nekrasov,  "Dynamics
Of Strings In Noncommutative Gauge Theory," hep-th/0007204,  JHEP
{\bf 0010:021} (2000).}. All this implies very interesting
interrelation between the non-abelian gauge degrees of freedom (of
open strings as the closest relatives of gauge theories in the
string world) and purely geometrical "gravitational"  degrees of
freedom (of closed strings). From the mathematical side it seems
to provide the most intriguing example of the unity of Algebra and
Geometry.

The simplest example of such relation is given by multiple
$D$-brane background. The important lesson learned from the recent
studies of the open string field theory is the new point of view on
such backgrounds. A few years ago one would think about $D$-branes
as solitons in the closed string theory, but by now it is
well-known that $D$-branes can be also considered as solitons in
the open string field theory. This can be seen both from cubic {\bf
CS} string field theory of \ref\witone{E. Witten, "Noncommutative
Geometry And String Field Theory," Nucl. Phys. {\bf B268} (1986),
235.} and background independent open string field theory
\ref\witbione{E. Witten, "On Background Independent Open String
Field Theory," hep-th/9208027, Phys. Rev. {\bf D46} (1992) 5467. },
\ref\witbitwo{E. Witten, "Some Computations In Background
Independent Open-String Field Theory," hep-th/9210065, Phys. Rev.
{\bf D47} (1993) 3405.}, \ref\sshone{S. Shatashvili, "Comment On
The Background Independent Open String Theory," hep-th/9303143,
Phys. Lett. {\bf B311} (1993) 83.}, \ref\sshtwo{S. Shatashvili, "On
The Problems With Background Independence In String Theory,"
Preprint IASSNS-HEP-93/66, hep-th/9311177, Algebra and Anal., {\bf
v. 6} (1994) 215. }. In the latter any boundary CFT (with
world-sheet being a disk) is a critical point for corresponding
string field theory action; this action evaluated on classical
solution coincides with the world-sheet partition function. Thus
any $D$-brane background should provide a classical solution for
the background independent open string field theory
lagrangian\foot{In fact this was already noticed in \witbitwo\ and
reinterpreted in modern terms in \ref\sh{S. Shatashvili,  "Closed
Strings As Solitons In Background Independent Open String Field
Theory," unpublished, talk at IHES, Paris, July 1997.} in order to
argue that field theory of open strings contains all possible
stringy backgrounds, including $D$-branes and closed strings, as
classical solutions and thus can serve as a definition of full
(consistent) off-shell string theory.}.

Although the above argument gives simple explanation of the fact that
$D$-branes are solitons in open string field theory, many issues
remain unclarified. For instance it is not clear how the
non-abelian gauge fields emerge in the case of the near coincident
$D$-branes in this formalism. The interpretation  of the closed
strings in these terms also remains an open question (see
discussion in lecture notes \ref\shindia{S. Shatashvili, "On Field
Theory Of Open Strings, Tachyon Condensation And Closed Strings,"
hep-th/0105076.} and comments below).

In this paper we will address the first question using the
formalism of background independent open string field theory
 which turned out to be
very effective in verifying the Sen's conjectures \ref\sen {A.
Sen,  "Stable Non-BPS States In String Theory," hep-th/9803194,
JHEP {\bf 9806:007} (1998); A. Sen  "Stable Non-BPS Bound States
Of BPS D-branes,"   hep-th/9805019, JHEP {\bf 9808:010} (1998) ;
A. Sen,  "Tachyon Condensation On The Brane Antibrane System,"
hep-th/9805170, JHEP {\bf 9808:012} (1998) ; A. Sen and B.
Zwiebach,  "Tachyon Condensation In String Field Theory,"
hep-th/9912249,  JHEP {\bf 0003:002} (2000).} as it was
demonstrated in \ref\gsone{ A. Gerasimov and S. Shatashvili, "On
The Exact Tachyon Potential In Open String Field Theory,"
hep-th/0009103,  JHEP {\bf 0010:034} (2000). }, \ref\kmm{D.
Kutasov, M. Marino and G. Moore,  "Some Exact Results On Tachyon
Condensation In String Field Theory,"  hep-th/0009148, JHEP {\bf
0010:045} (2000).}, \ref\gstwo{A. Gerasimov and S. Shatashvili,
"Stringy Higgs Mechanism And The Fate of Open Strings,"
hep-th/0011009,  JHEP {\bf 0101:019} (2001).}
 \foot{After this work
was finished the studies of similar questions in the lines of
cubic {\bf CS} string field theory appeared in \ref\gt{D. Gross
and W. Taylor,  "Split String Field Theory," hep-th/0105059.},
\ref\rsz{L. Rastelli, A. Sen and B. Zwiebach,  "Half Strings,
Projectors And Multiple D-branes In Vacuum String Field Theory,"
 hep-th/0105058; L. Rastelli, A. Sen and B. Zwiebach,  "Boundary
CFT Construction Of D-brane In Vacuum String Field Theory,"
 hep-th/0105168.}.}.

One of the main reasons for the appearance of the non-abelian
algebraic structures in the open string theory is an obvious
geometrical interpretation of the open string as a some kind of
matrix (open strings have two ends) acting in the appropriate
Hilbert space \ref\witone{E. Witten,  "Noncommutative Geometry And
String Field Theory,"  Nucl. Phys. {\bf B268} (1986) 253.},
\ref\wittwo{E. Witten,  "Some Remarks About String Field Theory,"
Phys. Scripta {\bf T15} (1987) 70.},  \ref\strominger{ A.
Strominger, "Closed Strings In Open String Field Theory," Phys.
Rev. Lett. {\bf v.58}  n.7 (1987) 629.}, \ref\astwo{G. T. Horowitz
and A. Strominger,  "Derivation And Associative Anomaly In Open
String Theory,"  Phys. Lett. {\bf B185} (1987) 45.},
\ref\asthree{G. T. Horowitz, J. Lykken, R. Rohm and A. Strominger,
"A Purely Cubic Action For  String Field Theory," Phys. Rev. Lett.
{\bf 57}  (1986) 283.}. The infinite dimensionality of this
Hilbert space leads to some unusual phenomena responsible for the
peculiar properties of string theory. Let us stress that the
non-commutativity is a basic property of string theory and  is
not a special feature of the backgrounds with non-zero vacuum
value of the $B$-field.

The existence of this "matrix" structure of the open string theory
allows us to add non-abelian indexes to the open string wave
function with the help of Chan-Paton factors. Basically one has a
tensor product of the stringy "matrix" algebra and some abstract
matrix algebra. One of the important outputs of the discovery of
the $D$-branes was the understanding that Chan-Paton factors also
have a geometrical origin and naturally
 appear in multiple $D$-branes backgrounds \Dbr.

 We argue that the non-commutative structure of the open string theory on a manifold
$M$ may be captured in a reasonable approximation by the
non-commutative algebra of differential operators on $M$. The
algebra of differential operators on the manifold together with
its various completions is the model of the matrix algebra
appearing in Matrix Theory of strings \BFSS. By considering the
algebra of differential operators instead of the (abstract)
infinite dimensional matrices of the Matrix Theory we partially
loose
 background independence but instead we get a very  explicit description
 of the non-commutative degrees of freedom.

The algebra of differential operators  may be considered as a
non-commutative deformation of the algebra of functions on the
cotangent bundle $T^*M$ defined by the canonical quantization with
the symplectic form $\omega=\sum_i dp_i \wedge dx^i$. The widely
discussed non-commutative structure appearing when $B$ -field is
non-zero \ref\CDS{A. Connes, M. R. Douglas and A. Schwarz,
"Noncommutative Geometry And Matrix Theory: Compactification On
Tori,"  hep-th/9711162, JHEP {\bf 9802:003} (1998). },
\ref\NScw{N. Nekrasov and A. Schwarz, "Instantons On
Noncommutative ${\IR}^4$, And (2,0) Superconformal Six Dimensional
Theory," hep-th/9802068, Commmun. Math. Phys. {\bf 198} (1998)
689.}, \ref\SW{N. Seiberg and E. Witten, "String Theory And
Noncommutative Geometry,"  hep-th/9912072, JHEP {\bf 9909:032}
(1999).}, from this point of view, may be described as  some
$B$-dependent deformation of the basic symplectic structure. Full
description of the non-commutative structure in the open string
theory obviously includes the differential operators on the space
of the maps of the (holomorphic) disks to the space-time. Some
remarks on this are given at the end of the paper.

We propose a construction of the vertex operators responsible for
Chan-Paton degrees of freedom. The main ingredients of this
construction turns out to be closely related to the algebra of
differential operators. It is not accidental. By looking at the
short distance correlation functions in the boundary conformal
field theory with appropriate regularization prescription we show
that they can be reproduced by a simple quantum mechanics
describing the quantization of the space which locally looks like
a cotangent bundle of the space-time where string theory is
defined. The corresponding algebra of the quantum operators
(differential operators acting on the space of functions on
space-time) is the basic structure one needs in order to extract
the matrix degrees of freedom entering in the description of the
$D$-brane theories. We give the explicit construction of this
reduction in terms of tachyon condensation for various tachyonic
profiles.

 We hope that our simple model for the description of the non-abelian degrees of
freedom of the open string theory, among other things could shed
some light on the open/closed strings correspondence in the lines
of discussion in \gstwo. As an illustration of the usefulness of
the approach  proposed in this paper, we apply it \ref\AGS{E.
Akhmedov, A. Gerasimov and S. Shatashvili,  "On Unification Of RR
Couplings," hep-th/0105228. } to the problem of the universal
description of the RR gauge field couplings with open strings.

One shall note that remarks regarding the importance of the
understanding of non-abelian degrees of freedom in terms of
boundary sigma models can be found in \ref\HKM{J. Harvey, D.
Kutasov and E. Martinec,  "On The Relevance Of Tachyon,"
hep-th/0003101.}, \kmm. The considerations in \ref\Harvey{J. A.
Harvey, P. Horava and P. Kraus,  "D-Sphalerons And The Topology Of
String Configuration Space,"  hep-th/0001143, JHEP {\bf 0003:021}
(2000). } might also be relevant to our discussion.

\newsec{Matrix algebra, Vertex operators and $GL(\infty)$}

In background independent open string field theory (we will
consider only the case when ghosts and matter decouple) the
classical action is defined on the space of boundary conditions
for bosonic string with world-sheet of disk topology. This action
has a critical point for any conformal boundary condition; also,
every critical point corresponds to a conformal boundary condition.
If the space of boundary conditions is parameterized by local
coordinates $t^i$ the classical action can be written in terms of
the $\beta$-function $\beta^i(t)$ corresponding to the coupling
$t^i$ and matter partition function $Z(t)$:
\eqn\betaeq{S(t)=-\beta^i(t)\pr_i Z(t) + Z(t)} On-shell ($t=t_c$ -
exactly marginal boundary perturbation) the action coincides with
the partition function $Z(t_c)$. If the world-sheet theory is
perturbed by the complete set of boundary operators (corresponding
to the full spectrum of given conformal field theory) the action
\betaeq\ is well-defined in the space of couplings although at
first glance the world-sheet theory might look non-renormalizable
\sshtwo. The simplest case where one can exactly compute the
effective action is given by the quadratic off-shell tachyon
profile \witbitwo: \eqn\tach{
 T(X(\theta))=T_0+\sum_{\mu}{1\over 2}u_{\mu} (X^{\mu}(\theta)-a^{\mu})^2}
(for the discussion of the limitations of this computation for
current application see
\shindia).

Since various collection of $D$-branes shall provide us with
world-sheet (disk) conformal field theories, the action \betaeq\
must have corresponding critical points. We shall show that
multi-branes are indeed the solutions of the equations of motion for
\betaeq. Besides, there shall be a vacuum without open strings
\sen\ - closed string vacuum. In terms of the profile, \tach\ the
closed string vacuum corresponds to the solution with
$T_0=\infty ,u_{\mu}=0$ of the equations of motion for $S$.
Solutions corresponding to $D$-branes can be represented by the
tachyonic profiles localizing the effective action on the
relevant submanifolds. For example, in order to explicitly write down
such solution one could use the approximation of the
delta function by the profile \tach\ with $T_0 \rightarrow \infty,
\quad
  u_1\rightarrow \infty, \quad u_{\mu \neq 1}=0$.
 This  corresponds to
  $D$-brane boundary condition which is described by the equation
  $X^1=a^1$. Let us consider decomposition of the fields on the boundary of the disk
  $X^{\mu}(\theta)=X^{\mu}+X^{\mu}_*(\theta)$ on the
  constant and non-constant parts, integrate out the
  non-constant modes and find the corresponding action in terms of the
  integral over $X$
(here we use integration by parts over $X$; for some details see
\shindia):

  \eqn\actionBI{S(T_0,u)=\int e^{-T(X)} F(\partial^2 T(X), \partial^3 T(X), ...) =
  \int dX e^{-\sum_{\mu}{1\over 2}u_{\mu}
  (X^{\mu}-a^{\mu})^2}F(T_0,u)}
One can solve the equations of motions for $T_0$ and evaluate the
action on such solution:
 \eqn\actionBIone{S(u)=\int dXe^{-\sum_{\mu}{1\over 2}u_{\mu}
  (X^{\mu}-a^{\mu})^2}\widetilde{F}(u)}
 Consider  the case when $u_{\mu \neq 1}=0$.
The action at the conformal point should be invariant with respect
to the scaling of $u_1$. This unambiguously fixes the asymptotics
of function $\widetilde{F}(u_1)$:
 \eqn\actionBItwo{S(u_1)\sim \int dX e^{-{1\over 2}u_1
  (X^1-a^1)^2}\sqrt{u_1}}
In the limit $u_1 \rightarrow \infty $ we have: \eqn
\actionBIthree{S(u) \rightarrow \int dX \delta(X^1-a^1)} Note the
correct normalization of the action is a consequence of the fact
that at the conformal point the action coincides with the
partition function. Thus we see the localization which is a
manifestation of the appearance of the $D$-brane. In terms of the
sigma model description this leads to the projector operator:
\eqn\sigmaone{ \int DX e^{\int d\theta (-uX^2(\theta))}\sim \int
DX DPe^{\int d\theta  ({1\over
u}P^2(\theta)+P(\theta)(X(\theta)-a))}\rightarrow_{u\rightarrow
\infty} \int DX\delta(X-a)}

For general tachyon potential an  arbitrary map of the disk to
the flat space-time can be written as a sum of harmonic map (with
boundary map being $X(\theta)$) and arbitrary function with zero
value on the boundary. Thus the world-sheet action perturbed by
the boundary operators (as in background independent open string
field theory) can be written purely in terms of the boundary value
$X(\theta)$ : \eqn\ws{\eqalign{& I_B= \int \int d\theta d \theta'
X^{\mu}(\theta) H(\theta, \theta') X_{\mu}(\theta')+\int d\theta
T(X(\theta)) \cr &= T(X) + \int X_*^{\mu}(\theta)
[H(\theta-\theta')\delta_{\mu \nu}+ \delta(\theta-\theta')
\partial_{\mu}\partial_{\nu} T(X)] X_*^{\nu} (\theta') +\cr
& \quad \quad \quad \quad
 + O(\partial^3 T(X))}}
here $ H(\theta , \theta') = {1\over 2} \sum_{k \in \IZ} e^{i k
(\theta - \theta')} |k| $.
 In the lowest, two derivative approximation one can ignore the last term
 and get for the world-sheet partition function:
 \eqn\part{
Z(T) = \int dX e^{-T(X)} det' [ H +
\partial^2 T]^{-{1 \over 2}}
}
This can be used for the computation of \betaeq\
(but only in two derivative approximation
as it was mentioned in \shindia; for higher derivative terms
the contributions of the last term in \ws\ mix with the
powers of $\partial^2 T$ coming from
determinant in \part).

Now consider the case of the multicritical tachyon profile, or in other words
- some kind of boundary Landau-Ginzburg theory with potential $W(X)^2$ (for
simplicity we will be treating only the case of non-trivial
dependence on the one coordinate of space-time - $X$):
\eqn\Tprofile{T(X(\theta))=T_0+u(W(X))^2 = T_0 + u \prod_{i=1}^N
(X-a_i)^2} At $u \rightarrow \infty $ (when the points $a_i$ are
far from each other) in this case we get the insertion of the
projector in the world-sheet path integral which defines the
space-time action \betaeq\ (from the superposition of the
solutions from different critical points of the tachyon profile)
and thus:

\eqn\actionmult{ S(u)\sim \sum_i\int dX e^{-{1\over 2}u T''(a_i)
  (X-a_i)^2}\sqrt{u T''(a_i)} \rightarrow \int dX (\sum_i
  \delta(X-a_i))}
  Note that the normalization in front of gaussian is uniquely fixed
  by considerations described above.

Thus, world-sheet path integral (and the string field theory
action ) in the boundary theory
 reduces to the trace over $n$-dimensional vector space as a result of above
 projection. This vector space may described as a subspace
(linear combination of the delta-functions)
 of the  linear space $\CF$ of all functions on the space-time.

Our goal will be to describe the operators corresponding to
"jumps" between the points of minima of the  potential $T(X)$.

 First we give a simple description in terms of the operators
acting on the space $\CF$. The operators we need
 should be compatible with the reduction \actionmult\
(they should preserve the linear space of the delta-functions
$\delta(X-a_i)$). Let us first note
that there are projection operators: \eqn\Eii{
 E_{ii}=\frac{\prod_{k\neq i}(X-a_k)}{\prod_{k\neq
 i}(a_i-a_k)}=\frac{W(X)}{W'(a_i)(X-a_i)}}
with the property: \eqn\mnaa{E_{ii}(X)\delta(X-a_j)=\delta_{ij}
\delta(X-a_j) }
\eqn\mnaaone{
E_{ii}(X)E_{jj}(X)=\delta_{ij}E_{ii}(X)  \, \quad \quad mod \quad
W(X)} In addition one can consider the set of differential
 (more exactly difference) operators:
\eqn\bnaa{
 E_{ij}=e^{(a_i-a_j)\pr_X} \frac{\prod_{k\neq j}(X-a_k)}{\prod_{k\neq
 j}(a_j-a_k)}=e^{(a_i-a_j)\pr_X}E_{jj}}
(in the special case $i=j$, $E_{ij}$ becomes $E_{ii}$ defined in
 \Eii).
 Their action does not spoil the structure defined  by reduction
\actionmult. It is easy to show that the following relation is
satisfied: \eqn\bnaaaa{
E_{ij}\delta(X-a_k)=\delta_{jk}\delta(X-a_i)} We conclude that
the
operators $E_{ij}$ modulo $W(X)$ generate the algebra $gl(N)$.

Now we would like to give an explicit realization of these
operators in terms of the operators of the corresponding boundary
conformal field theory. The non-abelian degrees of freedom have
their origin in the open strings stretching between $D$-branes. Let us
start with Dirichlet boundary condition $X_{|{\partial D}}=a$. It
is easy to construct the boundary operator in conformal field
theory that shifts the Dirichlet boundary conditions $X_{|{\partial
D}}=a$ to new boundary condition $X_{|{\partial D}}=a+\Delta
X(\theta)$: \eqn\two{V_{\Delta X}=e^{\int d\theta \pr_n X(\theta)
\Delta X(\theta)}} In order to realize the process when in the
beginning the string ends
 on the $D$-brane $X=a_i$ then jumps to $D$-brane $X=a_j$ and then at
 the end jumps back to the $D$-brane $X=a_i$ we take the
 "shift"- function  in the form of step-function: \eqn\jump{\Delta
 X=(a_1-a_2)\epsilon(\theta|\theta_1,\theta_2)}
 $$\epsilon(\theta|\theta_1,\theta_2)=1 \quad \quad \quad
\theta_1<\theta<\theta_2$$
 $$\epsilon(\theta|\theta_1,\theta_2)=0 \quad \quad \quad \quad
 {otherwise}$$
 Then using the decomposition
  of the scalar field $X=X_+ +X_-$ in terms of positive and negative
  frequency modes
\eqn\xfield{X^{\mu}(\theta)==X_+^{\mu}(\theta)+X_-^{\mu}(\theta)=
 (\frac{1}{2}X^{\mu}_0+\sum_{k >0}  X^{\mu}_k
 e^{ik\theta})+({1 \over 2}X^{\mu}_0+\sum_{k <0}  X^{\mu}_k
 e^{ik\theta})}
  we could represent
 $V_{\Delta X}$ operator as the product of two "jump"-operators:
 \eqn\three{V_{\Delta X}=e^{\int^{\theta_1}_{\theta_2}(a_i-a_j)\pr_n X}=
 e^{\int^{\theta_1}_{\theta_2}(\pr_{\theta}X_+ -
 \pr_{\theta}X_-)(a_i-a_j)}=}\eqn\threeone{
=e^{(X_+(\theta_1)-X_-(\theta_1))(a_i-a_j)}e^{-(X_+(\theta_2)-X_-(\theta_2))(a_i-a_j)}}
  Therefore one has the following expression for the "jump"-operator: \eqn\Vertex{V^{ij}(X(\theta))
   =e^{\frac{(a_i-a_j)}{\alpha'}(X_m(\theta))}, \quad \quad X_m = X_+-X_-}
  This operator corresponds to the excitation with the mass defined by its
  conformal dimension:
 \eqn\fourtwo{m^2=\mid a_i-a_j \mid ^2/(\alpha ')^2}
  which is in accordance with the expectations ($m^2\sim length^2$) for the mass of
the non-abelian part of the tachyon. In order to get the operators
describing the general states from the spectrum one should
multiply \Vertex\ by the polynomial of the derivatives of $X$ and
the standard exponential factors $e^{ipx}$ responsible for the
non-zero momentum along $D$-brane. One can make a contact with
\bnaa\ by realizing that the sector of the boundary theory
corresponding to the Dirichlet boundary condition $X=a_j$ is given
by the insertion of the $\delta(X-a_j)$. The action of $E_{ij}$ on
this function reproduces  the action of the operator $\exp
((a_i-a_j)\pr_X)$.

\newsec{Quantum analog of the Hamiltonian reduction
on $T^*X$ and Matrix algebra}

In this section  we explain the meaning of the operators $E_{ij}$
in a slightly different way. The following can be formulated  as
the quantum reduction of the algebra of quantized functions on
$T^*X$ \ref\Merk{S. Merkulov, "Deformation Quantization Of The
$n$-tuple Point,"   math.QA/9810158.} (this construction also appears
in the discussion of W-geometry \ref\Ger{A. Gerasimov,
unpublished, 1991.}).

Consider the simple case of one dimensional flat space $M$. The
differential operator may be considered as the function on the
non-commutative space $T^*M_{h}$ -the deformation of the cotangent
bundle defined by the canonical Poisson structure
$h\frac{\partial^2}{\partial x
\partial p}$. Let $D=\partial^n$ be a differential operator of
$n$-th order. The space of solutions is $Sol(D)=\{1,x,x^2,\cdots
,x^{n-1}\}$.

We would like to make quantum analog of the Hamiltonian reduction
with the momentum given by $D$. In other words we look at the
subspace $\CA$ of all differential operators with the following
properties. For any operator $P \in \CA$  there should exist an
operator $P'$ satisfying the condition: \eqn\defalgone{DP=P'D} and
we identify the operators in $\CA$ under the equivalence relation:
 \eqn\defalgoneone{ P \sim P+ P''D}
where $P''$ is an arbitrary operator. It is easy to see that the
operators from $\CA$ act on the space of solutions $Sol(D)$.
Moreover we have a multiplication  structure on $\CA$ (the product
of equivalence classes is unambiguously defined as the equivalence
class).
 Using the second condition we may restrict
ourselves by the operators of the order $(n-1)$:
$P=a_0+a_1\partial +\cdots
 +a_{n-1}\partial ^{n-1}$. From the first condition we get the
 restrictions on the coefficients $a_i$. Simple calculation shows
 that the number of free parameters is $n^2$.
It is not surprising that the $n^2$-dimensional algebra acting on
the $n$-dimensional space is isomorphic to $gl(n)$. For example,
in case of $n=2$ we have explicit isomorphism:
 \eqn\twodim{P=(p_0+p_1x)+(p_3+p_2x-p_1x^2)\partial}
\eqn\twodimis{P \sim
 \pmatrix{p_0 & p_1 \cr p_3 & (p_0+p_2)}}
 This is a standard representation of the $gl(2)$ algebra in terms
 of the first order operators.

  Consider  more interesting example:
$D=\partial^2-\lambda\partial$. The space of solutions
$Sol(D)=(1,e^{\lambda x})$ and the corresponding algebra is
generated by the operators: \eqn\algebralambda{P=(p_1 e^{\lambda
x}+p_0)+(-\frac{p_1}{\lambda}e^{\lambda x}+\frac{p_2}{
\lambda}e^{-\lambda x}+\frac{(p_3-p_0)}{\lambda})\partial_x}

One could make Fourier transformation and look at the ideal
defined by the function $W(x)=x^2-\lambda x$. The space of
solutions $f\in Sol(W)$ of the equation $W(X)f(X)=0$ is given by
delta-functions $Sol(W(x))=\{\delta(x),\delta(x-\lambda)\}$ and
using the Fourier transform of \algebralambda\ we get:
\eqn\alg{P=\frac{1}{\lambda}(-p_1 e^{-\lambda
\pr_x}(x-\lambda)+p_2e^{\lambda \pr_x}x+p_3-p_0(x-\lambda))} The
action of these operators on the delta-function bases of $Sol(W)$
provides the connection with \Vertex . These considerations could
be  straightforwardly generalized to the case of an arbitrary
(multidimensional) potential $W(x)$.

\newsec{Algebra of the differential operators from the boundary field
theory}

In the last section we will show how the framework of the
qunatization of the cotangent bundle of the space-time naturally
appears in the sigma model description of open strings.

As a model example we consider bosonic sigma-model on the disk
with  unspecified boundary conditions:
 \eqn\action{S_{2D}=\int d\theta d\tau
 ||dX||^2}
The action takes extreme values on the configurations:
 \eqn\eomone{\Delta X=0}
 \eqn\eomtwo{\partial_n  X|_{{\partial D}}=0}
where $\partial_n$ is the normal derivative on the boundary. Thus
on the classical solutions Neumann boundary conditions hold.
Boundary action was given in \ws: \eqn\First{I_B= \int \int
d\theta d \theta' X_{\mu}(\theta) H(\theta,\theta')
X^{\mu}(\theta')}
 Here $H=d*$ is a
normal derivative operator acting on the harmonic functions on the
disk. The meaning of this notation is the following. Introduce the
operation $*$ on the fields ("T-duality"):
\eqn\xmfield{*X^{\mu}(\theta)=\sum_{k \in \IZ} sign(k)\, X^{\mu}_k
e^{ik\theta}= X_+^{\mu}(\theta)-X_-^{\mu}(\theta)} (compare with
\xfield) $d*$ may be considered as the composition of the usual
derivative and $*$-operator.
 Note that at least when the local operators
are considered the fields $*X$ and $X$ may be treated as
independent. Two-point correlation functions  of the fields are
given by:
 \eqn\corfunone{<X^{\mu}(\theta)X^{\nu}(\theta')>=\delta^{\mu \nu}\sum_{k\neq
 0} \frac{1}{|k|}e^{ik(\theta-\theta')}}
 \eqn\corfuntwo{<*X^{\mu}(\theta)*X^{\nu}(\theta')>=\delta^{\mu \nu}\sum_{k\neq
 0} \frac{1}{|k|}e^{ik(\theta-\theta')}}

 \eqn\corfunthree{<X^{\mu}(\theta)*X^{\nu}(\theta')>=\delta^{\mu \nu}\sum_{k\neq
 0} \frac{1}{
 k}e^{ik(\theta-\theta')}}

The non-symmetric part  of these correlators  leads to
non-commutativity of the operators $*X$ and $X$:
 \eqn\comutone{[X^{\mu}(\theta),X^{\nu}(\theta')]=0}
 \eqn\comuttwo{[*X^{\mu}(\theta),*X^{\nu}(\theta')]=0}
 \eqn\comutthree{[*X^{\mu}(\theta),X^{\nu}(\theta')]\sim
 \delta^{\mu \nu}\delta(\theta-\theta')}

The symmetric part of the correlation functions is given by the
standard Green function: \eqn\eqone{G(\theta,\theta')=2\log
|e^{i\theta}-e^{i\theta'}|^2} and is singular when the points
coincide. This leads to the necessity to regularize the theory.
 Let us suppose that our regularization prescription
modifies the metric on the boundary as follows:
\eqn\eqtwo{
 ||z_1-z_2||=|z_1-z_2| \quad \quad \quad when \quad \quad
\quad \quad \quad \quad \quad |z_1-z_2|\gg l_R}
 \eqn\eqtwoo{
 ||z_1-z_2||=1 \quad \quad \quad when \quad \quad \quad |z_1-z_2|\ll l_R }
(thus $l_R$ is a characteristic regularization length). With this
prescription we get for modified Green function $G_R(0)=0$. Note
that this regularization does not modify the "non-commutative"
part of the correlators. Thus we would like to conclude that if we
are interested in the small distance correlation functions only
"non-commutative" parts survive. One should propose that this is
the way matrix degrees of freedom show up in the sigma model
approach. It would be interesting to extract relevant degrees of
freedom explicitly. This can be done as follows. Let us start with
the action: \eqn\eqthree{
 I_B=\int d\theta X^{\mu}d*X^{\mu}}
 It is obvious that this theory is equivalent to the following
 theory with the additional fields $X^m_{\mu}$:
 \eqn\lagrone{ I_B'=\int d\theta(X^m_{\mu} dX^{\mu}+ X^m_{\mu}d*X^m_{\mu})}

To have equivalence between the theories \eqthree\ and \lagrone\
one should not integrate over zero modes of the new fields
$X^m_{\mu}$ in the functional integral (it may be achieved by
insertion of the delta function $\delta(X^m_{\mu}(0))$).

Now consider a simple example of the calculation of the
correlation function in this theory: \eqn\eqonea{
 <\prod_i e^{ip_iX(\theta_i)}>_{I_B'}}
It is clear  that when the points $\theta_i$ are sufficiently
close correlation functions at small distances are dominated by
the first term in the lagrangian \lagrone\ (with the
regularization \eqtwo ,\eqtwoo). Thus we are getting the effective
theory: \eqn\eqtwoa{
 I=\int d\theta (X^m_{\mu} dX^{\mu})}
This is the standard description of the quantization of the
cotangent bundle. The basic operators have the commutation
relations:\eqn\eqthreea{ [X^m_{\mu},X^m_{\nu}]=0}
\eqn\eqthreebb{[X^{\mu},X^{\nu}]=0}\eqn\eqoneaa{
[X^m_{\mu},X^{\nu}]=\delta_{\mu}^{\nu}}

Let us show that the correlation functions of the operators
\three\ and \Vertex\  reduce to the  standard representation of
the differential operators  in the theory with the action
\lagrone. Consider the correlation function that includes normal
derivatives: \eqn\examone{
 <\prod_i e^{ip_iX(\theta_i)} e^{\int d\theta
 a_{\mu}(\theta)d*X^{\mu}(\theta)}>_{I_B'}}

 One could get rid of the non-local terms by the shift of the
 variable. For
 the correlation function \examone\ the simple shift: \eqn\app{X^m_{\mu}\rightarrow
X^m_{\mu}+*a_{\mu}} makes the necessary change. We are interested
 in  the short distance behaviour of the
correlation functions. Thus after dropping the terms with
non-localities with
* operations (this is legitimate with our regularization
prescription), we come to the following correlation function in the
theory with the action \lagrone\ : \eqn\arr{ <\prod_i
e^{ip_iX(\theta_i)} e^{\int d\theta
 a^{\mu}(\theta)dX^m_{\mu}(\theta)}>_{I_B'}}

Now let us take the function $a$ of the form:
$a(\theta)=(a_1-a_2)\epsilon(\theta|\theta_1,\theta_2)$ (compare
with \jump). We get a standard path integral representation of the
product of differential operators $e^{\pm (a_1-a_2)\pr_X}$ and
$\prod_i e^{ip_iX}$ in quantum mechanics. Thus we have shown that
the effective theory is consistent with description of the
non-abelian degrees of freedom discussed in the first and second
sections of the paper.

\newsec{Conclusions and further directions}

In this paper we have discussed the appearance of non-abelian
degrees of freedom in open string theory using background
independent open string field theory. The importance of the
algebra of differential operators for the description of this
phenomena was stressed. This algebraic structure may be helpful
for the construction of the full effective action for tachyon
field \shindia. This action may be connected with natural
"geometric" action for the algebra of differential operators
analogous to the action functionals on coadjoint orbits and
related symplectic manifolds considered in \ref\AFS{A. Alekseev,
L. Faddeev and S. Shatashvili,  "Quantization Of Symplectic Orbits
Of Compact Lie Groups By Means Of The Functional Integral," J.
Geom. Phys. 5:391-406  (1989);  A. Alekseev and S. Shatashvili,
  "From Geometric Quantization To Conformal Field Theory,"
 Commun. Math. Phys. {\bf 128} (1990) 197.}.

Obviously there is interesting mathematics behind the
descriptions of the vector bundles on the submanifolds in terms of
differential operators.  We give a very  elementary description
and do not discuss the connections with D-modules, Derived
categories (see for instance \ref\doug{M. Douglas,  "D-branes,
Categories And  N=1 supersymmetry,"  hep-th/0011017.}) and its
K-(co)homology invariants.  The application to the universal
construction of the  couplings of RR gauge fields with open
strings in  an arbitrary $D$-brane background  is given in \AGS.

Let us make at the end several remarks on the generalization of
 this approach to the full-fledged description of all string
 modes. The description in terms of the differential operators
 acting on the functions on the space-time  obviously is an
 approximation of the more general structure of differential
 operators on the space of the half-strings (open string field theory)
 \witone ,  \wittwo ,   \strominger , \astwo , \asthree .
Consider an arbitrary open string as a composition of the
"infinitesimal" open strings with identified  ends. Each
"infinitesimal" open string may be described by the local
variables and taking into account the considerations in this paper
one could suggest that the description in terms of the
differential operators may be a natural candidate. Then stringy
"matrix" algebra would be substituted  by (subalgebra of) the
universal enveloping algebra of differential operators as an
abstract Lie algebra. However, the real state of the matters is
more complicated. In order to deal with arbitrary functionals on the
configuration space of the open strings, these "infinitesimal"
strings should contain the information about all derivatives of the
map of the small interval into the space-time. From the Lagrangian
point of view this leads to the consideration of the boundary
interaction containing arbitrary derivatives of the fields (see
related discussion in \gstwo, \shindia. In the
Hamiltonian language this is beyond the quantization of
a tangent bundle  of space - time. One
could believe that generalization of the differential operators
appearing in this more general setup would provide the right
framework for the discussion of string algebra.  This
picture seems to be rather close to the description of the
strings in Matrix Theory and deserves further considerations.

 {\bf Acknowledgements:} We would like to thank E. Akhmedov, M. Douglas, I. Frenkel, Hong Liu,
 A. Morozov, N. Nekrasov, L. Takhtajan, E. Verlinde and E. Witten. S. Sh.
also would like to thank Rutgers New High Energy Theory Center for hospitality.
  The research of A. G. was partially supported  by  Grant for Support of
Scientific Schools 00-15-96557 and by RFBR 00-02-16530 and the
research of S. Sh. is supported by OJI award from DOE.

\listrefs

\bye